\documentclass[prl, twocolumn]{revtex4-1}
\usepackage{amsmath}
\usepackage{amssymb}
\usepackage{graphicx}
\usepackage{color}
\usepackage{multirow}

\newcommand{\be}{\begin{equation}}
\newcommand{\ee}{\end{equation}}
\newcommand{\bea}{\begin{eqnarray}}
\newcommand{\eea}{\end{eqnarray}}
\newcommand{\balg}{\begin{align}}
\newcommand{\ealg}{\end{align}}

\def\rrscan{\ensuremath{{\mathrm{r}{^2}\mathrm{SCAN}}}}

\begin{document}
  \title{Meta-GGA Performance in Solids at Almost GGA Cost}
  \author{Daniel Mej{\'i}a-Rodr{\'i}guez}
  \email{dmejiarodriguez@ufl.edu}
  \author{S.B.\ Trickey}
  \email{trickey@qtp.ufl.edu}
  \affiliation{Center for Molecular Magnetic Quantum Materials,
Quantum Theory Project, Department of Physics, University of Florida,
Gainesville, FL 32611}
  \date{27 Aug. 2020}
	
\begin{abstract}
  \noindent A recent modification, $\rrscan$,  of the SCAN
  (strongly constrained and
  appropriately normed) meta-GGA exchange-correlation functional 
  mostly eliminates numerical instabilities and attendant integration grid
  sensitivities exhibited by SCAN.
  Here we show that the successful deorbitalization of SCAN to SCAN-L (SCAN with
  density Laplacian dependence) carries over directly to yield $\rrscan$-L.
  A major benefit is that the high iteration counts that hindered use of 
  SCAN-L are eliminated in $\rrscan$-L. It therefore is a computationally much faster
  meta-GGA than its orbital-dependent antecedent.  Validation data
  for molecular heats of formation, bond lengths, and vibration frequencies (G3/99X, T96-R, T82-F test sets respectively)
  and on lattice constants, and cohesive energies (for 55 solids)
  and  bulk moduli (for 40 solids) are provided.  In addition, we show that the over-magnetization of
  bcc Fe from SCAN persists in $\rrscan$ but does not appear in
  $\rrscan$-L, just as with SCAN-L.
\end{abstract}

  \maketitle

\textit{Setting and Motivation} - Recognition of chemically
significant electron density inhomogeneities by use of an 
indicator function (usually denoted $\alpha$; see below) is the critical
mechanism by which a meta-GGA 
exchange-correlation (XC) functional improves upon a generalized
gradient approximation (GGA). The most successful meta-GGA so far (see
\cite{IsaacsWolverton2018} and references therein), is SCAN, the
strongly constrained and appropriately normed functional
\cite{SCAN,SCANNature}.  Its success is attributed to 
enforcement upon it of all known rigorous constraints which a meta-GGA
can meet, together with calibration to the energies of certain
primitive physical systems (the ``appropriate norms''; see
Supplemental Material to Ref. \cite{SCAN}).

Despite its successes, SCAN introduced a numerical problem and
exacerbated a methodological challenge.  We defer discussion of the
methodological issue briefly.  The numerical problem has two
elements. SCAN exhibits high sensitivity to numerical integration grid
density. Handling that requires extremely dense, hence costly, grids.  The other
element is instability of self-consistent field convergence that is
hard to foresee, hence control, for a given system (especially in periodic
solids).

Those two numerical issues with SCAN were 
addressed by  Bart\'ok and Yates \cite{rSCAN} by a
simple renormalization of the denominator of $\alpha$, a rescaling, and
replacement of the SCAN switching function $f(\alpha)$ with a smoother
seventh-degree polynomial for $\alpha < 2.5$. The resulting
revised SCAN (rSCAN) is
far better behaved computationally than SCAN. rSCAN preserves both
the good molecular bond lengths and vibrational frequencies performance
of SCAN. Unfortunately, rSCAN does not preserve
the good performance of SCAN for benchmark molecular heats of formation
\cite{rSCANcomment}.  In periodic solids, SCAN and rSCAN are about the
same for lattice constants and cohesive energies 
\cite{rSCANcomment} on a 55 solid test set \cite{SCAN+rVV10}
and for bulk moduli on a 44 solid set \cite{TranStelzlBlaha}.

Very recently Furness et al. \cite{FurnessEtAl2020} have shown that
the shortcomings of rSCAN stem from the fact that its regularization
resulted in violation of several constraints satisfied by SCAN.
They adopted the smooth switching function strategy of rSCAN 
combined with modifications to restore compliance with all but one of the
constraints satisfied in SCAN.  The result, their regularized-restored
SCAN functional or $\rrscan$, recovers the strong performance trends
of SCAN relative to molecular and solid data sets while maintaining
the numerical stability of rSCAN.

The $\rrscan$ combination of accuracy and stability opens an opportunity for
equally improved response to the methodological challenge.  That comes  
from the regularized chemical region detector 
\be
\alpha(\mathbf{r}):= \frac{\tau_s -\tau_W}{\tau_{TF} + \eta \tau_W} \; .
\label{alphadefn}
\ee
Here $\tau_s = (1/2)\sum f_j \vert \nabla
\varphi_j(\textbf{r})\vert^2$ is the positive-definite Kohn-Sham (KS)
kinetic energy density in terms of the KS orbitals $\varphi_j$ and
occupations $f_j$, $\tau_W$ and $\tau_{TF}$ are the von
Weizs\"acker and Thomas-Fermi kinetic energy densities respectively,
and $\eta = 10^{-3}$ is a small regularization parameter. The original
$\alpha$ has $\eta =0$.
Obviously the orbital dependence of the XC energy introduced by $\alpha$
disqualifies SCAN or $\rrscan$ from being used in orbital-free DFT.
Worse, that orbital-dependence makes an ordinary KS
calculation almost prohibitively costly because it necessitates
an optimized effective potential calculation 
\cite{StadeleMajewskiVoglGoerling1997,GraboKreibichGross1997,GraboKreibachKurthGross2000,HesselmannGoerling2008}
at every SCF step. Usual practice to evade that cost is to do
generalized-KS (gKS) calculations. The gKS Euler equation  
follows from variation of the density functional approximation with
respect to the orbitals, not the density.  For meta-GGA and higher-rung
functionals  the KS and gKS equations are
not equivalent \cite{YangPengSunPerdewGKSBandGap2016,PerdewEtAlBandGaps2017}.

We addressed this challenge by deorbitalization \cite{SCANL1,SCANL2,Tran2018}.  
The deorbitalized version of SCAN, SCAN-L, differs from SCAN only in
using an approximate orbital-independent
approximation for $\tau_s$ in the original $\alpha$ to
give $\alpha [n, \nabla n, \nabla^2n]$ 
(with $n$ the electron number density).
Deorbitalization restores use of the KS equation. Furthermore, a SCAN-L
calculation  should be much faster than SCAN. In practice that advantage
oft-times
went unrealized because numerical instabilities caused very slow
SCF convergence. Experience\cite{CancioWagner}
suggested that the problem might be rooted in the 
$\nabla^2 n$ dependence. By deorbitalizing $\rrscan$ into $\rrscan$-L, we
show here that much of the problem actually was inherited from SCAN.  

\textit{Procedure and Results} -

The deorbitalization of $\rrscan$ used precisely the same
deorbitalized $\tau_s$ form and parametrization as was used for
SCAN-L in Refs. \cite{SCANL1,SCANL2}.   Molecular calculations
were done with a locally modified developers' version of the
NWChem code \cite{NWChem}. Similarly, the periodic solid
calculations were done with a locally modified version of VASP 5.4.4 \cite{vasp3,vasppaw}.
Note the remarks about coding in the VASP meta-GGA trunk in Ref.
\cite{SCANL2}.  As in that reference, we did calculations with
coding implemented in that trunk (to check unambiguously against
the VASP results of Ref. \cite{FurnessEtAl2020}) and coding in
the GGA trunk (to ascertain optimal speed-up from the deorbitalization).
Basis sets, cutoffs, and other matters of technique were as
documented in \cite{SCANL1,SCANL2} with one exception, the
PAWs, documented below.
  
Table \ref{REFnwchem} summarizes the results for the molecular test
sets in the form of mean 
absolute errors (MAEs) with respect to experiment for heats of formation 
(G3/99X set \cite{G3}), bond lengths (T96-R set \cite{t96rt82f}), 
and harmonic vibrational frequencies (T82-F set \cite{t96rt82f}) obtained with  
the NWChem \texttt{HUGE} grid. For lower-density grids, Table \ref{REFnwchem} 
shows the mean absolute deviation (MAD) and maximum absolute deviation (MAX)
with respect to the \texttt{HUGE} grid results. 

\begin{table*}
\centering
\caption{Performance of SCAN, \rrscan, and \rrscan-L for heats of formation (kcal/mol), bond lengths ({\AA}), and vibrational frequencies (cm$^{-1}$) 
at various grid densities. Mean absolute errors with respect to experiment
from the NWChem \texttt{HUGE} grid calculations are in \textbf{boldface}. 
For the lower-density grids, mean absolute deviation and maximum absolute deviation 
(in parenthesis), with respect to those \texttt{HUGE} results are shown.
\label{REFnwchem}}
  \begin{tabular}{l l c c c c c}
       &                                & SCAN            & SCAN-L          &      & $\rrscan$    & $\rrscan$-L  \\\toprule
\multicolumn{2}{l}{Heats of formation}  & \bf 4.93        & \bf 5.66        &      & \bf 4.49     & \bf 5.30     \\\hline 
  \phantom{111} & \texttt{coarse}       & 5.92 (26.61)    & 5.12 (22.02)    &      & 0.40 (1.80)  & 0.81 (5.90)  \\
       & \texttt{medium}                & 2.31 (15.56)    & 2.36 (14.63)    &      & 0.09 (0.54)  & 0.22 (1.19)  \\
       & \texttt{fine}                  & 0.73 {\ }(4.25) & 0.88 {\ }(4.59) &      & 0.01 (0.09)  & 0.04 (0.22)  \\
       & \texttt{xfine}                 & 0.23 {\ }(1.42) & 0.36 {\ }(1.90) &      & 0.00 (0.02)  & 0.01 (0.07)  \\\hline
\multicolumn{2}{l}{Bond lengths}        & \bf 0.009       & \bf 0.011       &      & \bf 0.010      & \bf 0.011         \\\hline
       & \texttt{coarse}                & 0.001 (0.016)   & 0.001 (0.014)   &      & 0.000 (0.003)  &   0.001 (0.006)  \\
       & \texttt{medium}                & 0.001 (0.006)   & 0.001 (0.006)   &      & 0.000 (0.002)  &   0.000 (0.001)  \\
       & \texttt{fine}                  & 0.000 (0.004)   & 0.000 (0.004)   &      & 0.000 (0.000)  &   0.000 (0.001)  \\
       & \texttt{xfine}                 & 0.000 (0.004)   & 0.000 (0.003)   &      & 0.000 (0.000)  &   0.000 (0.001)  \\\hline
\multicolumn{2}{l}{Vib. frequencies}    & \bf 31.1        & \bf 28.8        &      & \bf 30.9       & \bf 25.6       \\\hline
       & \texttt{coarse}                & 24.2 (150.5)    & 24.1 (183.0)    &      &  7.5 (71.2)    & 7.0  (55.9)    \\
       & \texttt{medium}                & 18.4 (330.2)    & 21.7 (156.4)    &      &  2.1 (21.1)    & 2.1  (22.0)    \\
       & \texttt{fine}                  & 10.5 (100.0)    & 14.8 (130.5)    &      &  1.3 (11.6)    & 1.4  (12.9)    \\
       & \texttt{xfine}                 &  3.6 (32.1)     &  5.1 {\ }(39.9) &      &  0.5  (3.9)    & 0.6   (4.2)     \\\toprule
  \end{tabular}
\end{table*}

Table \ref{REFnwchem} confirms the necessity of very dense numerical
integration grids for both SCAN and SCAN-L. Even the \texttt{XFINE}
preset grid yields deviations above 1 kcal/mol and 30 cm$^{-1}$
(bond lengths are less problematic). In contrast, \rrscan\ and
\rrscan-L results are well-converged with the \texttt{MEDIUM} and
\texttt{FINE} grid presets. This is a major improvement both in
reliability as well as in performance, since numerical integration
easily can become a computational bottleneck. See below regarding
calculation of $\nabla^2 n$ on the integration grid in the context
of a Gaussian-type orbital (GTO) basis.

It is important to note
the disentanglement of instabilities due to the functional form versus 
the Laplacian-dependence. It now is clear that SCAN-L exhibits roughly
the same stability difficulties as SCAN because their common
structure. However, the highly stable \rrscan-L shows thermochemical deviations
about three times larger than those for \rrscan\ on a given preset grid. This
residual grid sensitivity seems directly attributable to the
Laplacian-dependence of \rrscan-L.

Different from the setup described in \cite{SCANL2}, the periodic solid 
calculations in VASP used the "no-suffix" PAW datasets instead of the
GW ones.  (Discussion of this choice is below.) Since the PAW datasets used here are softer than the GW sets, we lowered the
kinetic energy cutoff to 600 eV. The k-point sampling density was 
increased to match that reported in \cite{FurnessEtAl2020} by using
the \texttt{KSPACING=0.1} keyword.

Table \ref{refVASP} shows MAEs with respect to experimental
results for three crystalline test sets \cite{SCANL2}: 55 equilibrium
lattice constants (with cubic
or hexagonal symmetry), 40 bulk moduli (cubic symmetry), and 55
cohesive energies. Zero-point effects were removed from 
experimental lattice parameters and bulk moduli. 

Similar to the molecular case, Table \ref{refVASP} shows 
that deorbitalization of \rrscan\ is achieved with the
same success as with SCAN. (Note that the values in that table are not directly
comparable with our previous reports \cite{SCANL2,rSCANcomment} because of 
the PAW dataset change.) In both cases, lattice parameters
are well-treated. Predictions of bulk moduli and cohesive energies with both
deorbitalized functionals show large percentage deviations, 
but the deviation magnitudes are nonetheless quite small (large percentage
error in a small quantity).  Some physics also is involved.  Part of
the cohesive energy MAE difference between $\rrscan$-L and $\rrscan$
comes from the different electronic configurations found for the
W atom. $\rrscan$-L, SCAN, and SCAN-L all  find an 6s$^1$5d$^5$ valence
when the configuration is 
unconstrained, while $\rrscan$ finds the correct 6s$^2$5d$^4$ configuration.

\begin{table*}
\caption{Mean absolute error comparison for SCAN, SCAN-L, $\rrscan$, and $\rrscan$-L XC functionals
for the solid state test sets. \label{refVASP}}
\begin{tabular}{l c c c c c}
                                 & SCAN  & SCAN-L   & & \rrscan & \rrscan-L \\\toprule
Lattice parameters [\AA]         & 0.034 & 0.038   & & 0.037    & 0.039   \\
Bulk moduli [GPa]                & 7.4   & 8.8     & & 6.0      & 8.9     \\
Cohesive energies [eV/atom]      & 0.21  & 0.21    & & 0.20     & 0.33     \\\toprule
\end{tabular}
\end{table*}

The total time needed to converge the 660 single-point 
calculations (12 per each solid) was used as a surrogate measure
of the speed and stability of each functional. 
Consistent with expectations, the total times relative to the SCAN benchmark
were 0.924 for $\rrscan$, 0.438 for SCAN-L, 0.272 for $\rrscan$-L, and 0.227 for
PBE \cite{PBE}. In other words, $\rrscan$-L computational cost in a plane-wave
basis is almost as inexpensive as
a standard GGA functional, even though numerical demands  
associated with the Laplacian-dependence remain.

There is an important caveat. The SCF stability of all the
approximate functionals, as measured by the number of iterations
needed for convergence, can be degraded by use of the GW
PAW datasets.  The Laplacian-dependent functionals are significantly
worse in this regard than the orbital-dependent ones.  What one sees
with some, but not all, of the GW datasets is erratic SCF
convergence.  In those cases, near-SCF-convergence
from iteration steps $N-1$ to $N$ often is followed by drastic
worsening at step $N+1$.  Exploration of the origins of this behavior
is outside the scope of the present work.

Despite the fact that SCAN-L inherits many properties from SCAN,
SCAN-L avoids the over-magnetization of simple elemental solids 
such as bcc Fe \cite{OverMag},  We therefore tested \rrscan\ and
\rrscan-L in bcc Fe at the experimental
lattice constant (2.86 \AA). As with SCAN and rSCAN, \rrscan\ predicts
the magnetic moment of bcc Fe to be 2.63 $\mu_B$/atom. \rrscan-L
lowers that to 
2.27 $\mu_B$/atom, in line with other GGA functionals and SCAN-L.

\textit{Summary} - The slow (sometimes extremely so) SCF convergence
and numerical sensitivities of SCAN-L originate mostly in
structural characteristics of SCAN. The removal of those provided by
$\rrscan$ leads to a similarly well-behaved deorbitalized version,
$\rrscan$-L.  Except for the elemental 3d solid magnetization
discrepancy, $\rrscan$-L replicates the behavior of $\rrscan$ on major
molecular and solid benchmarks. $\rrscan$-L additionally provides a
pure KS treatment (hence enables band-gap and optical excitation
comparison with gKS results from $\rrscan$),  and should, in most
cases support significantly faster solid calculations than $\rrscan$, on the
time scale of an ordinary GGA.  Further speedup of molecular
calculations from $\rrscan$-L in a GTO basis
will require addressing the remaining computational bottleneck,
calculation of $\nabla^2 n$ from GTO second-derivatives on the
integration grid (rather than directly as in a plane-wave code.)
Nonetheless calculations with $\rrscan$-L with the \texttt{MEDIUM}
NWChem grid are as fast as those with SCAN with the \texttt{XFINE}
grid.

\begin{acknowledgments}
This work was supported by 
U.S.\ Dept. of Energy Energy Frontier Research Center 
grant DE-SC 0019330.
\end{acknowledgments}

\section{Supplemental Material}
Detailed molecular test set results  
obtained with $\rrscan$ and $\rrscan$-L, as well as
the detailed results for the periodic solid
test sets for all functionals, can be 
found in the supplementary material \cite{SuppMat}.


\begin{thebibliography}{99}

  \bibitem{IsaacsWolverton2018}
  E. B. Isaacs and C. Wolverton,
  Phys. Rev. Mat.\ \textbf{2}, 063801 (2018).

  \bibitem{SCAN}
  J. Sun, A. Ruzsinszky, and J.P. Perdew,
  Phys. Rev. Lett.\ \textbf{115}, 036402 (2015).

  \bibitem{SCANNature}
  J.\ Sun, R.C.\ Remsing, Y.\ Zhang, Z.\ Sun, A.\ Ruzsinszky, H.\ Peng,
  Z.\ Yang, A.\ Paul, U.\ Waghmare, X.\ Wu, M.L.\ Klein, and J.P.\ Perdew,
  Nature Chemistry \textbf{8}, %831--836 (2016).
  831 (2016).

\bibitem{rSCAN}
A. P. Bart\'ok and J. R. Yates,
J.\ Chem.\ Phys.\ \textbf{150}, 161101 (2019).

\bibitem{rSCANcomment}
  D. Mej\'ia-Rodr\'iguez and S.B.\ Trickey,
  J. Chem. Phys. \textbf{151}, 207101 (2019)].
  
\bibitem{SCAN+rVV10}
H.\ Peng, Z.-H.\ Yang, J.P.\ Perdew and J.\ Sun, Phys.\ Rev.\ X \textbf{6}, 041005 (2016).

\bibitem{TranStelzlBlaha}
F.\ Tran, J.\ Stelzl and P.\ Blaha, J.\ Chem.\ Phys.\ \textbf{144}, 204120 (2016).

\bibitem{FurnessEtAl2020}
%''Accurate and numerically efficient r$^2$SCAN meta-generalized gradient approximation'',
  James W. Furness, Aaron D. Kaplan, Jinliang Ning, John P. Perdew, and
  Jianwei Sun, arXiv 2008.03374v1, 11 Aug. 2020.
    
\bibitem{StadeleMajewskiVoglGoerling1997}
%``Exact Kohn-Sham Exchange Potential in Semiconductors''
M.\ St\"adele, J.A.\ Majewski, P.\ Vogl and A.\ G\"orling,
Phys.\ Rev.\ Lett.\ \textbf{79}, 2089 (1997)

\bibitem{GraboKreibichGross1997}
%``Optimized Effective Potential for Atoms and Molecules'', 
T.\ Grabo, T.\ Kreibich and E.K.U.\ Gross, 
Molec. Eng. \textbf{7}, 27 (1997).

\bibitem{GraboKreibachKurthGross2000} 
%``Orbital Functionals in Density Functional Theory: the Optimized 
%Effective Potential Method'',  
T.~Grabo, T.~Kreibach, S.~Kurth, and E.K.U.~Gross, 
in {\it Strong Coulomb Correlations in Electronic Structure: Beyond 
the Local Density Approximation}, V.I.~Anisimov ed.\ (Gordon and 
Breach, Tokyo, 2000) 203. 
 
\bibitem{HesselmannGoerling2008}
%``Comparison between optimized effective potential and Kohn-Sham methods'',   
A.\ He{\ss}elmann and A.\ G\"orling, 
Chem.\ Phys.\ Lett.\ \textbf{455}, 110 (2008) and refs.\ therein.   

\bibitem{YangPengSunPerdewGKSBandGap2016}
%``More realistic band gaps from meta-generalized gradient 
%approximations: Only in a generalized Kohn-Sham scheme'',  
Z.H.\ Yang, H.\ Peng, J.\ Sun, and J.P.\ Perdew, 
Phys.\ Rev.\ B \textbf{93}, 205205  (2016) 

\bibitem{PerdewEtAlBandGaps2017} 
J.P.\ Perdew, W.\ Yang, K.\ Burke, Z.\ Yang, E.K.U.\ Gross, 
M.\ Scheffler, G.E.\ Scuseria, T.M.\ Henderson, I.Y.\ Zhang, 
A.\ Ruzsinszky, H.\ Peng, J.\ Sun, E.\ Trushin, and A.\ G\"orling,
Proc.\ Nat.\ Acad.\ Sci.\ (USA) \textbf{114}, 2801 (2017).

  \bibitem{SCANL1}
  D. Mej\'ia-Rodr\'iguez and S.B.\ Trickey,
  Phys. Rev. A\ \textbf{96}, 052512 (2017).

  \bibitem{SCANL2}
  D. Mej\'ia-Rodr\'iguez and S.B.\ Trickey,
  Phys. Rev. B\ \textbf{98}, 115161 (2018).
   
  \bibitem{Tran2018}
  F. Tran, P. Kovacs, L. Kalantari, G.K.H. Madsen, and P. Blaha,
  J. Chem. Phys.\ \textbf{149}, 144105 (2018).

\bibitem{CancioWagner}
``Laplacian-based generalized gradient approximations for the exchange energy'',
Antonio C. Cancio and Chris E. Wagner, arXiv 1308.3744
  
\bibitem{NWChem}
  %''NWChem: Past, present, and future'',
  
  E. Apr{\`a}, E.J. Bylaska, W.A. de Jong, N. Govind, K. Kowalski,
  T.P. Straatsma, M. Valiev, H.J.J. van Dam, Y. Alexeev, J. Anchell,
  V. Anisimov, F.W. Aquino, R. Atta-Fynn, J. Autschbach, N.P. Bauman,
  J.C. Becca, D.E. Bernholdt, K. Bhaskaran-Nair, S. Bogatko, P. Borowski,
  J. Boschen, J. Brabec,  A. Bruner, E. Cau{\"e}t, Y. Chen, G.N. Chuev,
  C.J. Cramer, J. Daily, M.J.O. Deegan, T.H. Dunning Jr., M. Dupuis,
  K. G. Dyall, G.I. Fann, S.A. Fischer, A. Fonari, H. Fr{\'u}chtl,
  L. Gagliardi, J. Garza, N.  Gawande, S. Ghosh, K. Glaesemann, A. W. G{\"o}tz,
  J. Hammond, V. Helms, E.D. Hermes, K. Hirao, S. Hirata,  M. Jacquelin,
  L. Jensen, B.G. Johnson, H. J{\'o}nsson, R.A. Kendall, M. Klemm, R.
  Kobayashi, V. Konkov,  S. Krishnamoorthy, M. Krishnan, Z. Lin, R.D. Lins,
  R.J. Littlefield, A.J. Logsdail, K. Lopata, W. Ma, A.V. Marenich,
  J. Mart{\'i}n del Campo, D. Mej{\'i}a Rodr{\'i}guez, J.E. Moore, J.M. Mullin,
  T. Nakajima, D.R. Nascimento, J.A. Nichols, P.J. Nichols, J. Nieplocha,
  A. Otero-de-la-Roza, B. Palmer, A. Panyala, T. Pirojsirikul, B. Peng,
  R. Peverati, J. Pittner, L. Pollack,  R.M. Richard, P. Sadayappan, G.C.
  Schatz, W.A. Shelton, D.W. Silverstein, D.M.A. Smith, T.A. Soares, D. Song,
  M. Swart, H.L. Taylor, G. S. Thomas, V. Tipparaju, D.G. Truhlar, K.
  Tsemekhman, T. Van Voorhis, {\'A}. V{\'a}zquez-Mayagoitia, P. Verma,
  O. Villa, A. Vishnu, K.D. Vogiatzis, D. Wang, J.H. Weare, M.J. Williamson,
  T.L. Windus, K. Wolinski, A.T. Wong, Q. Wu, C. Yang, Q. Yu, M. Zacharias,
  Z. Zhang, Y. Zhao, and R.J. Harrison,
  J. Chem. Phys. \textbf{152}, 184102 (2020)  

  \bibitem{vasp3}
  G. Kresse and J. Furthm\"uller, Phys.\ Rev.\ B\ \textbf{54}, 11169 (1996).

  \bibitem{vasppaw}
  G. Kresse and D. Joubert, Phys.\ Rev.\ B\ \textbf{59}, 1758 (1999).

\bibitem{G3}
L.A.~Curtiss, K.~Raghavachari, P.C.~Redfern, and J.A.~Pople, J.\ Chem.\ Phys.\ \textbf{106}, 1063 (1997); 
L.A.~Curtiss, P.C.~Redfern, K.~Raghavachari, and J.A.~Pople, J.\ Chem.\ Phys.\ \textbf{114}, 108 (2001). 

\bibitem{t96rt82f}
V.N.~Staroverov, G.E.~Scuseria, J.~Tao, and J.P~Perdew, J.\ Chem.\ Phys.\ \textbf{119} 12129 (2003);
V.N.~Staroverov, G.E.~Scuseria, J.~Tao, and J.P~Perdew, J.\ Chem.\ Phys.\ \textbf{121} 11507 (2004).

\bibitem{PBE}
J.P. Perdew, K. Burke, and M.~Ernzerhof,  
Phys. Rev. Lett. \textbf{77}, 3865 (1996); erratum
\textit{ibid.} \textbf{78}, 1396 (1997). 

\bibitem{OverMag}
D. Mej{\'i}a-Rodr{\'i}guez, and S.B. Trickey,
Phys. Rev. B \textbf{100}, 041113(R) (2019).

 \bibitem{SuppMat} Citation to Supplemental Material here.
  \end{thebibliography}
\end{document}